\documentclass[vecphys]{svmult}

\newcommand{\be}{\begin{equation}}
\newcommand{\ee}{\end{equation}}
\newcommand{\bea}{\begin{eqnarray}}
\newcommand{\eea}{\end{eqnarray}}
\newcommand{\bd}{\begin{displaymath}}
\newcommand{\ed}{\end{displaymath}}
\newcommand{\ej}[1]{e^{#1 ik}}

\usepackage{makeidx}         
\usepackage{graphicx}        
\usepackage{multicol}        
\usepackage[bottom]{footmisc}
\usepackage{dcolumn}
\usepackage{bm}
\usepackage{mathrsfs}
\usepackage{amsmath}
\usepackage{amssymb}
\usepackage{amsfonts}
\usepackage{wasysym}
\usepackage{mathbbol}
\DeclareMathAlphabet{\mathpzc}{OT1}{pzc}{m}{it}

\makeindex             

\begin{document}

\title*{Molecular Conductance from Ab Initio Calculations: 
       Self Energies and Absorbing Boundary Conditions}
\titlerunning{Self Energies and Absorbing Boundary Conditions}
\author{F. Evers\inst{1}\and
A. Arnold\inst{2}}
\institute{Institute of Nanotechnology, Research Center Karlsruhe,
  76021 Karlsruhe, Germany
\texttt{Evers@int.fzk.de}
\and Institut f\"ur Theorie der Kondensierten Materie,
 Universit\"at Karlsruhe, 76128 Karlsruhe, Germany }
%
%
\maketitle

\begin{abstract}
Calculating an exact self energy for ab initio
transport calculations relevant to {\em Molecular Electronics}
can be troublesome.
Errors or insufficient approximations
made at this step are often the reason why many
molecular transport studies become inconclusive.
We propose a simple and efficient approximation scheme, that
follows from interpreting the self energy as
an absorbing boundary condition of an
effective Schr{\"o}dinger equation. 
In order to explain the basic idea,
a broad introduction into the
physics incorporated in these self energies is given.
The method is further illustrated using a tight binding wire as
a toy model. Finally, also more realistic applications
for transport calculations based on the density functional theory
are included.
\end{abstract}

\section{Introduction}

The most important driving force in the research field of
{\em Molecular Electronics} are prospects on
technological applications -- whence the name --
in entirely new realms of system
parameters
\cite{likharev03}. 
The development of these new technologies also requires
serious progress in several disciplines
of fundamental sciences including both,
theory and experiment.
One of the major
theoretical challenges is the quantitative description of
transport through a molecule with a given contact geometry
\cite{reimers03,evers03prb,kurth:035308,sai05,burke05a}.

In order to appreciate the caliber of the problem, recall that
describing transport requires to keep track of two aspects of
physical reality, each by itself posing a task of considerable
difficulty. Needed are 
a) a good know\-ledge of molecular states, i.~e.
energy levels and orbitals, 
which is not easy to obtain, since they experience a strong influence
by Coulomb interactions on the
molecule, and 
b) a thorough understanding of 
the hybridization of these orbitals with the 
electronic lead states, so as to predict the broadening,
i.~e. the ``life time'', of molecular energy
levels. This seriously complicates matters for ab initio
calculations, because inevitably a macroscopic
number of degrees of freedom is involved.
We are facing here a classical dilemma:
a single one of the two problems 
-- interactions on the molecule and the macroscopic number of lead atoms --
 by itself can be dealt with reasonably well, 
but only at the expense of applying 
methods that exclude a simultaneous solution of 
the other problem. 
In a sense, we find ourselves in a situation not unlike
Ulysses, when he was trying to pass by
Scylla and Charybdis \cite{homer}. 

In this paper we present a method, that simplifies b),
i.~e. including 
macroscopic electrodes into ab initio calculations.
The incarnation, that we put forward
in this communication, 
operates in those instances where the calculation of 
transport coefficients builds upon a formalism in terms
of Green's functions. The basic idea developed here,
however, is much more general than that
and may also be of use for example in
transport calculations based on the density matrix
renormalization group \cite{bohr05}. 

A typical example of a Green's function based transport theory
met in cases, where the quasi-particles are
effectively non-interacting, is the Landauer-B{\"u}ttiker approach to
transport \cite{landauer57,buettiker85}.
In it the conductance (in units $e^2/h$)
is expressed as a transmission at the Fermi energy, 
$g{=}T(E_{\rm F})$. Explicitly, $T(E)$ has a representation
\be
T(E) = \text{tr}\  G \Gamma_{\mathpzc{l}} G^\dagger \Gamma_{\mathpzc r}
\label{ea}
\ee
which may be derived using elementary scattering theory
\cite{xue01,brandbyge02,evers03prb},
the Keldysh technique \cite{meir92} or the Kubo formula
\cite{fisher81,baranger89}, in principle. 

The ``dressed'' Green's functions, $G$, required in any of these approaches
describe the propagation of particles with energy $E$
on the molecule in the presence of the electrodes. The external leads,
left and right, enter these functions
by self energy contributions,
$\Sigma_{\mathpzc{l,r}}$, one for every electrode.
They relate $G$ to
the Green's function of the isolated molecule, $G_\mathpzc{M}$  by 
\be
  G^{-1}(E) = G^{-1}_{\mathpzc M}(E) - \Sigma_{\mathpzc{M}}(E)
  \label{e1}
  \ee
and include all the effects of coupling to the left and right leads,
$\Sigma_{\mathpzc{M}}{=}\Sigma_{{\mathpzc l}}{+}\Sigma_{{\mathpzc r}}$.
Also, they determine the level broadening $\Gamma_{\mathpzc
  l,r}{=}i(\Sigma_{\mathpzc l,r}{-}\Sigma^\dagger_{\mathpzc l,r})$
appearing in (\ref{e1}).
( Equation (\ref{e1}) should be understood as a family of
matrix equations with resolvent operators $G, G_{\mathpzc M}$,
parameterized by energy, $E$; Green's functions
are actually the matrix elements of these operators,
  $G(E,{\bf x},{\bf x'}) {=} \langle {\bf x} | G(E) | {\bf x'}
  \rangle$ and 
  $G_{\mathpzc M}(E,{\bf x},{\bf x'}) {=} \langle {\bf x} |
  G_{\mathpzc M} (E) | {\bf x'} \rangle$.
)

The calculation of the exact couplings, $\Sigma_{\mathpzc l,r}$, usually
is fairly troublesome. In the simplest case, when the electron interaction
can be appropriately dealt with by an effective single particle model,
the couplings take a structure
\be
       \Sigma_{{\mathpzc x}} = t_{\mathpzc x} G_{\mathpzc{S};{\mathpzc x}} t^\dagger_{\mathpzc x},
       \qquad {\mathpzc x}={\mathpzc l,r}
\label{e1b}
\ee
where $t_{\mathpzc l,r}$ denote the two hopping matrices
that connect the
molecular junction with the left and right electrodes
\cite{meir92}.
The ``surface'' Green's function, $G_{\mathpzc{S};{\mathpzc l,r}}$,
debuting here describes the propagation of quasi-particles on the
electrodes in the presence of the contact surface.
Even in this situation
calculating $\Sigma_\mathpzc{M}$ is not really easy.
Complications arise since 
a) $\Sigma_\mathpzc{M}$  should include {\em macroscopic}
leads plus  contact geometry and
b) the hopping matrix $t$ is (normally)
not just of the nearest neighbor type, so the contact
surface also involves sub-surface layers of
electrode atoms, in general.

The procedure to be proposed in this communication
simplifies the conductance calculation by essentially
eliminating the step of evaluating $\Sigma_{\mathpzc{M}}$.
It works after the molecule has been redefined.
The ``extended'' molecule, $e\mathpzc{M}$,
not only includes the original molecule but also pieces
of the left and right electrodes:
\be
G^{-1}(E) = G_{e\mathpzc{M}}^{-1} - \Sigma_{e\mathpzc{M}}.
\label{e1c}
\ee
Our key observation is the following: while the molecular conductance
is crucially dependent on microscopic details incorporated in
$\Sigma_{\mathpzc{M}}$, it is completely indifferent towards
details in $\Sigma_{e\mathpzc{M}}$, {\em if} $e\mathpzc{M}$ includes
sufficiently many electrode atoms.
As a consequence, there is no need to use the exact self energy
$\Sigma_{e\mathpzc{M}}$ in order to obtain (in principle) exact
results.
\footnote{The words ``exact result'' have in the present context
  a restricted meaning: they refer to the exact solution of the
  single particle scattering problem, that can be stated once the
  Kohn-Sham orbitals and energies are given. Under which conditions
  -- if at all -- scattering theory based on (ground state)
  Kohn-Sham orbitals could give an exact description of the full many body
  problem, this is an important question which, however, goes well beyond the
  scope of the present article.}
One can replace it by a simplistic model coupling of the type
\be
    \langle {\bf x}| \ \Sigma_{e\mathpzc{M}}(E)\  |{\bf x'}\rangle
    \to i \eta({\bf x})  \ \delta({\bf x}-{\bf x'}),
\label{e6}
\ee
where we have introduced a ``local leakage function'' $\eta({\bf x})$. 
It is crucial to our method, that fine tuning 
$\eta({\bf x})$ is obsolete
once certain criteria to be specified in
Sec. \ref{s:IIC} are satisfied.

The outline of this paper is as follows.
In Sec. \ref{s:II} we recapitulate
in broad terms the physical effects,
that are encoded in the self energy
formalism. The concept of the extended molecule
will emerge quite naturally from these
considerations. They will also illuminate under what
conditions (\ref{e6}) can be justified. 
In the following Secs. \ref{s:III} and \ref{s:IV} we
will present a series of model problems.
In order to demonstrate the principle, we begin in
Sec. \ref{s:III} with a
tight binding chain, for which numerical results
can be compared against analytical solutions. 
To illustrate the usefulness in practically relevant cases,
the conductance of di-thiophenyl is investigated in
Sec. \ref{s:IV} using an approach to transport based on the density
functional theory and the quantum chemistry package
TURBOMOLE \cite{turbomole,turbomole-ridft,turbomole-aux}.
This study is also intended to reveal the limits of the
ansatz (\ref{e6}). 


\section{\label{s:II} Basic Physics of Self Energies}

In this section we will give a more precise definition and a
justification of the procedure (\ref{e6}) for
constructing a self energy, which is based on  physical arguments.
In order to explain the logic, we first
recall on a qualitative level,
which physical information is carried by the original self energy,
$\Sigma_{\mathpzc{M}}$. We shall illustrate then, how this information is
transfered into $G_{e\mathpzc{M}}$
by reformulating the problem  to calculate $G$
in terms of an extended molecule. If enough metal atoms have been
included in $e\mathpzc{M}$, the ``information transfer''
will be complete.  Then the remaining information in the self energy
$\Sigma_{e\mathpzc{M}}$ of $e\mathpzc{M}$ is trivial, i.~e. it is no longer
molecule specific. Therefore, $\Sigma_{e\mathpzc{M}}$
is apt to simple approximations like (\ref{e6}).

\subsection{Self Energy of the Molecule $\Sigma_{\mathpzc{M}}$}

The self energy, $\Sigma_{\mathpzc{M}}$,
that appears in (\ref{e1}),
$$
  \Sigma_{\mathpzc{M}}(E)= G^{-1}_{\mathpzc M}(E) - G^{-1}(E) 
$$
has two qualitatively different effects which are incorporated into
its hermitian and anti-hermitian constituents.

\subsubsection{Hermitian Constituent}
The eigenvalues of $G_{e\mathpzc{M}}$ for the isolated molecule are
real numbers. Due to the hermitian piece,
$\delta H_\mathpzc{M}{=}(\Sigma_{\mathpzc{M}}{+}\Sigma^{\dagger}_{\mathpzc{M}})/2$,
these eigenvalues undergo a shift, $\Delta \epsilon_{\nu}$,
when the molecule is coupled to the electrodes.
In the case of weak electron-electron interaction this simple
``renormalization'' of excitation energies is all that can happen.
However, if the interaction is strong,
such that the electrons are highly correlated,
additional and qualitatively different effects can occur.
A most prominent representative is the Kondo effect, observation of
which has been reported in various recent experiments \cite{park02,liang02,heersche05}.
It manifests itself
in the spectral function of the coupled molecule
\be
\label{e6a}
    A(E) = (i/2\pi) \ \mathrm{Tr}(G(E)-G^\dagger(E)),
\ee
which measures the number of molecular excitations
with a certain energy, roughly speaking \cite{abrikosov63}.
Kondo-physics is signalized  
by an additional peak in $A(E)$,
the ``Abrikosov-Suhl''-resonance,
which sits right at the Fermi-energy of the leads \cite{mahan}.
This resonance is a collective
phenomenon involving electrons from the leads and the molecule;
it cannot be understood as a renormalization of a molecular
energy level alone.

Even in the absence of strong correlation effects, the
shift of molecular excitation levels, $\Delta \epsilon_{\nu}$, can have
very important consequences for the interpretation of experimental
findings. The presence of the metal electrodes can help screening
the interaction of electrons on the molecule. As a consequence, the
energy difference between the lowest unoccupied molecular energy
level (LUMO) and the highest occupied level (HOMO) will generally
shrink. In the extreme case, where the LUMO falls below
(or the HOMO above) the Fermi energy of the electrodes,
charge will flow onto the molecule such that the molecular
junction becomes partially polarized even at equilibrium
conditions.

\subsubsection{\label{s:IIA2}Anti-Hermitian Constituent:
  Exponential Time Evolution and Reser\-voirs}

The hermitian piece of $\Sigma_{\mathpzc{M}}$
is basically ``just'' a modification of the (effective)
Hamiltonian. By contrast, the anti-hermitian piece of the self energy,
$(\Sigma_\mathpzc{M}-\Sigma^\dagger_\mathpzc{M})/2$,
introduces a qualitatively new aspect, because
it gives rise to an imaginary component, $i\gamma_{\nu}$,
of the eigenvalues of $G$.
It gives the molecular levels, $\nu$, a finite lifetime
reflecting a simple physical fact: an initial excitation,
localized  at time $t{=}0$ on the molecule, can fade away to be  
absorbed by the leads, ultimately.


Let us discuss how excitations pass away in more detail so as to see,
why the electrodes and the thermodynamic limit are important
ingredients in understanding the self energy.
We begin by noting, that Green's functions can describe a time evolution of the
physical system. Therefore,  the relaxation rates, $\gamma_\nu$, also have a
straightforward interpretation in time space.
Assume, that the molecular junction
is prepared in an initial state such that the molecule has an excess
charge. Then, the  rates $\gamma_\nu$ describe an exponential
decay in time, $\exp(-\gamma_\nu t)$,  exhibited by
each contribution to this charge made from a certain molecular
level, $\nu$.

Now, the exponential dependence exposed here
is implied to be valid at all times including 
in particular the asymptotic regime $t\to \infty$.
This means, that the charge is really swallowed by the electrodes, 
it never returns to the molecule and only for this reason
the relaxation process can ever become complete.
In other words, the electrodes act
like  thermodynamic baths or {\em reservoirs}.
They destroy information about the
initial state in the sense that the return time
of a signal, i.~e. electrons,
from the reservoirs is infinite. 

As usual, a truly diverging time scale can be realized only 
with infinitely many degrees of freedom;
otherwise return paths (e.~g. of electrons) exist
with an overall weight that is not vanishing. 
In this infinite dimensional Hilbert space the time evolution is
unitary, of course. The (anti-hermitian part of the)
self energy pops up as a consequence of
projecting the full time evolution down to the
subspace of the molecule, $\mathpzc{M}$,
which then can no longer remain unitary. 
The principle encountered here is well known in
the general theory of non-equilibrium phenomena \cite{brenig89}.

\subsubsection{Self Energy and Transport}

Clearly, the decay rates $\gamma_\nu$
must be closely related to transport
properties, because they govern the time evolution of
charge exchange between molecules and leads.
Note, however, that the self energy
of the Green's function contains
information only about the total loss rate,
$$
\Sigma_{\mathpzc{M}}= \Sigma_{\mathpzc l} + \Sigma_{\mathpzc r}
$$
due to leakage. It does not necessarily keep track of the
rates $\Sigma_\mathpzc{l,r}$ separately, that describe the exchange with the
individual leads, left or right. 
This latter piece of information is important
for the transport characteristics, as can be seen e.g. in the
Landauer-B{\"u}ttiker formula (\ref{ea}). In general,
it cannot be reconstructed from the $G(E,{\bf x},{\bf x'})$ alone,
without making further assumptions (e.g. that
$\langle{\rm x}|\Sigma_{\mathpzc{M}}|{\bf x'}\rangle$
is block-diagonal with the two diagonal
entries resembling$\Sigma_{\mathpzc{r,l}}$,  separately).

In order to illustrate 
the significance of the level shifts $\Delta \epsilon_{\nu}$
and level broadenings $\gamma_{\nu}$ for the transport problem,
we consider now a situation typical of experiments on molecular conductance.
We focus on the case of weakly interacting electrons and 
call $\delta_{\rm hl}$ the energy gap between the HOMO
of the isolated molecule, $\epsilon_{h}$
and its LUMO, $\epsilon_{l}$:
$\delta_{\rm hl} {=}\epsilon_{l}{-}\epsilon_{h}$.
In typical transport experiments one has a situation, where
$\delta_{\rm hl}\apprge 1 {\text eV}$. At the same time,
the experimentally measured values of the conductance, $g$,
of the molecule only very rarely exceed $0.1$. 
Both observations taken together give a strong indication that for this
type of experiments the level broadening of HOMO and LUMO,
$\gamma_{\rm h,l}$,
is well below the level separation, $\gamma_{\rm h,l} \ll \delta_{\rm hl}$.
Roughly speaking, the conductance (\ref{ea}) will then be given
by a superposition of two Lorenzians, 
\be
    g = \sum_{{\rm x}={\rm h,l}} \frac{\gamma_{{\rm
        x},\mathpzc{l}}\gamma_{{\rm
        x},\mathpzc{r}}}{(E_{\rm F}-\epsilon_{\rm x}
      - \Delta\epsilon_{\rm x})^2 +
      (\gamma_{{\rm x},\mathpzc{l}} + \gamma_{{\rm x},\mathpzc{r}})^2/4}.
\label{e6b}
\ee
with a Fermi energy of the metal, $E_{\rm F}$, situated in between the
values of HOMO and LUMO after coupling,
$\epsilon_{\rm H,L}=\epsilon_{\rm h,l}+\Delta\epsilon_{\rm h,l}$.

We add a remark regarding uncertainties in theoretical
predictions of level positions and their broadenings.
Inaccuracies in calculating absolute values of the level positions
tend to induce a shift of the transmission curve,
but do not normally change their structure --
unless molecular levels happen to cross the Fermi energy
of the electrodes, of course. Often, the shift is
very similar for all energy levels involved, and therefore it can be
partially eliminated when the transmission is plotted over $E-E_{\rm F}$. 

Inaccuracies in the level broadening are more severe, 
since their error turns out to be of the order of unity.
The value of the
conductivity off resonance is determined by $\gamma^2_{\rm h,l}$,
and so a quantitative calculation of $g$ under these conditions
is very difficult. The source of this error and the question
how it can be overcome became a very active field of
research, recently
\cite{reimers03,evers03prb,kurth:035308,sai05,burke05a}.

\subsubsection{\label{s:IIA4}Relation to the Renormalization Group Method}

In this section, we describe the physics incorporated in
$\Sigma_{\mathpzc{M}}$ from the point of view of an hypothetical 
renormalization group method. This is to say,
that we investigate how $\Sigma_\mathpzc{M}$ 
evolves when we build up the molecular junction gradually step by step,
attaching more and more electrode atoms.
The idea is in a spirit  similar to the 
density matrix renormalization group \cite{schollwock05}.
The flow thus induced will be smooth unless the molecule
becomes strongly distorted, which could signalize for example
dissociation or ionization.  

In order to illustrate this evolutionary process, 
we
have performed calculations based on the density functional theory
(DFT) using the standard functional BP86 \cite{perdew86b,becke88}.
DFT provides us with an effective single particle Hamiltonian,
$H_{N_\mathpzc{E}}$,
with eigenvalues $\epsilon_{\nu}$ and corresponding eigenstates
$|\nu\rangle$.
Our interest is in how the eigenvalues and eigenfunctions
change when we include a gradually increasing number of
electrode atoms, $N_\mathpzc{E}$, in our model system. 

\begin{figure}[htbp]
  \centering
  \includegraphics[width=1.0\linewidth]{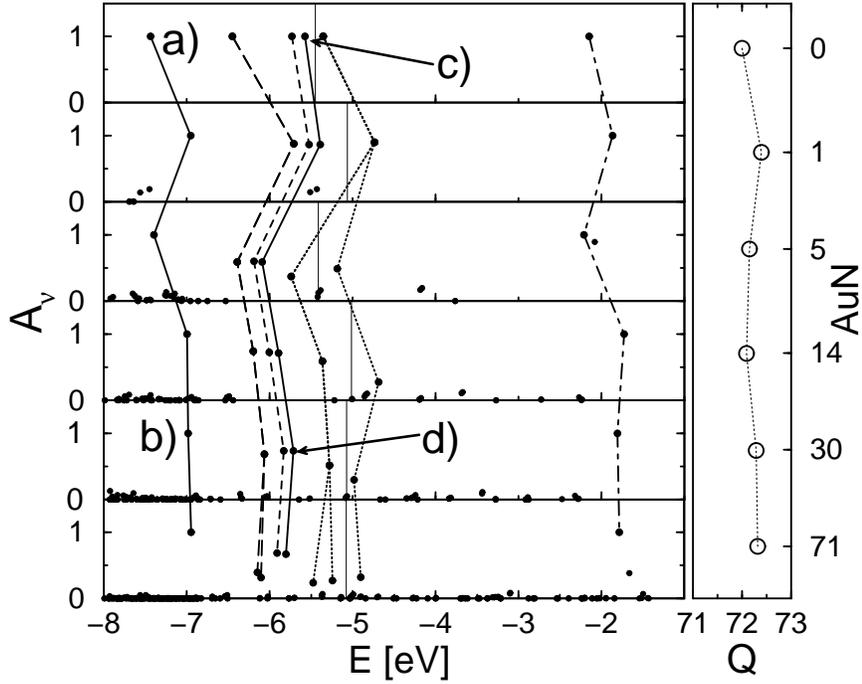}

  \caption{Di-thiophenyl between Au-pyramids (symmetric coupling,
    see Fig. \ref{f5a}). Left panel: flow of 
  energy of different orbitals with increasing
  number, $N_{\mathpzc{E}}$, of electrode Au-atoms as indicated
  at the right hand side axis. (Orbitals at a),b),c),d) are
  shown in Fig. \ref{f1orb}.) 
  Each orbital is characterized by its energy and weight $A_{\nu}$
  on the molecule.
  The vertical bars near $-5$eV mark the (center of the)
  HOMO-LUMO gap. The evolution of the six
  orbitals of the isolated molecule has been indicated by
  ``world'' lines. Right panel: evolution of charge, $Q$,
  accumulated on the molecule (including sulphor atoms).}
  \label{f1b}
\end{figure}

\begin{figure}
\begin{center}
  \includegraphics[width=0.5\linewidth]{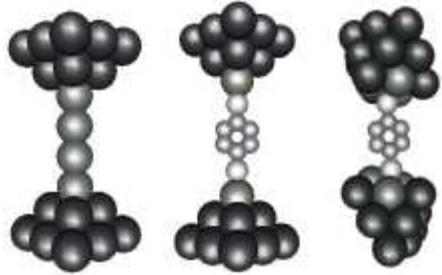}%
  \end{center}
\caption{\label{f5a} Atomic configurations of molecular junctions
  attached to pieces of the electrodes.
  Electrodes are modeled by pyramids, $N_{\mathpzc{E}}{=}14$
  Au-atoms each.
  Dark atoms are surface atoms used
  in self energy (\ref{e24}). 
  Left: 2 Au-atom wire, $N_{\mathpzc{S}}{=}13$ (Sec. \ref{ss:IIIA}).
  Center: di-thiophenyl with (stretched) ${\rm S-Au}_1$ coupling; 2 sulphor, six carbon
  atoms, $N_{\mathpzc{S}}{=}13$ (Sec. \ref{sss:IIIB1}).
  Right: di-thiophenyl with relaxed ${\rm S-Au}_3$ coupling,
  $N_{\mathpzc{S}}{=}11$ (Sec. \ref{sss:IIIB2}).}
\end{figure}

The result of this procedure has been depicted in Fig. \ref{f1b}
for the case of the molecule di-thiophenyl
(See Fig. \ref{f5a} for the detailed atomic structure.)
Every eigenfunction is represented by a data point
$(\epsilon_{\nu},A_{\nu})$.
The integrated amplitude is defined as
$$
A_{\nu} = \mathrm{Tr}_{\mathpzc{M}} |\nu\rangle\langle\nu|
$$
where the $\mathrm{Tr}_\mathpzc{M}$ is over the projected
segment of the Hilbert space
that is associated with the molecular degrees of freedom.
Our calculation is performed using a local
basis set $|X\ell\rangle$ (TZVPP \cite{schaefer94}),
with basis functions labeled by atomic
positions, $X$,  and orbital quantum numbers, $\ell$.
When evaluated in this basis,
the $\mathrm{Tr}_{\mathpzc{M}}$ is a sum over all basis states that
belong to molecular (i.~e. non-Au) atoms.

\begin{figure}[htbp]
  \centering
\begin{tabular}{cc}
  \includegraphics[height=2.5cm]{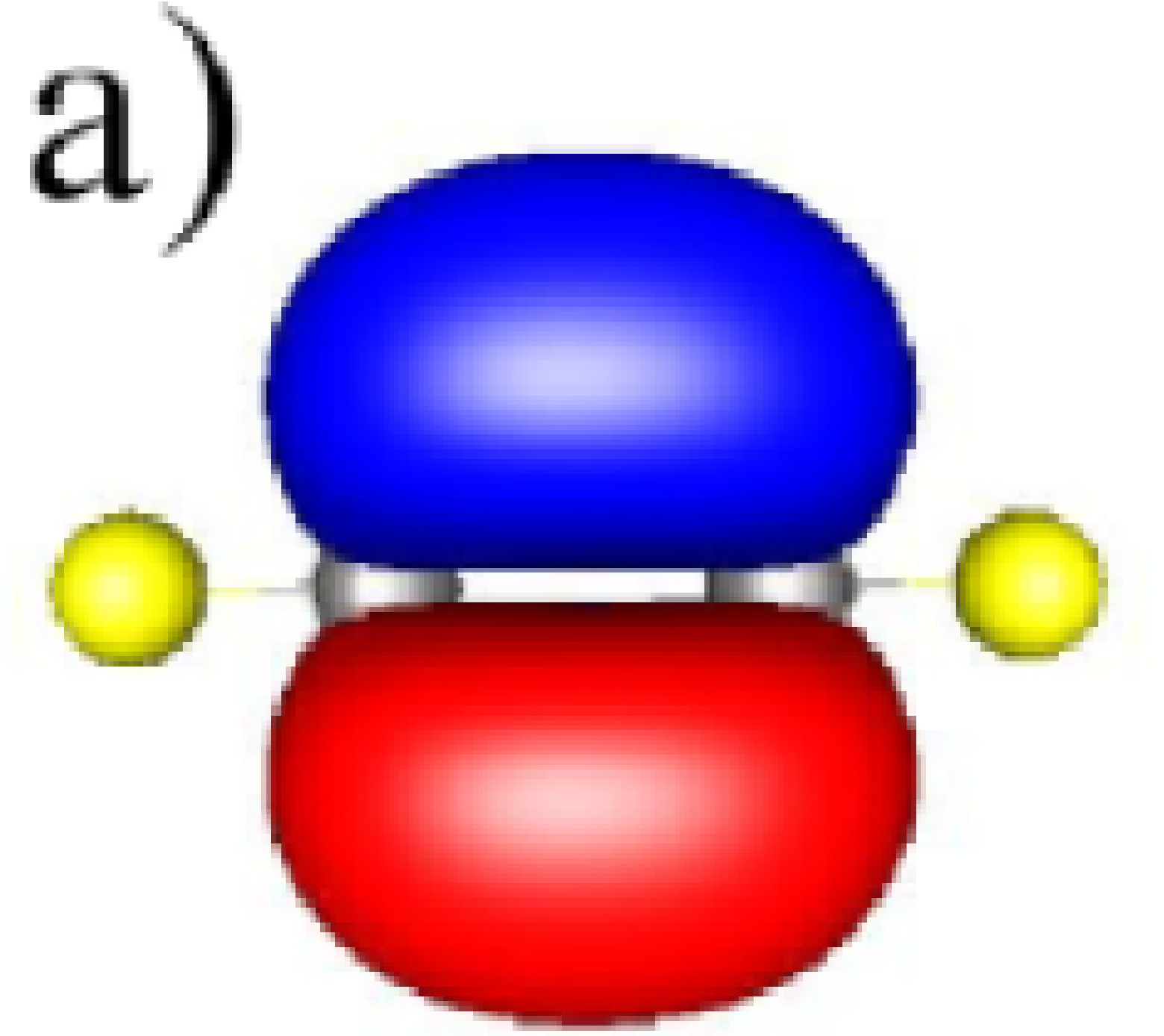}&
  \includegraphics[height=2.5cm]{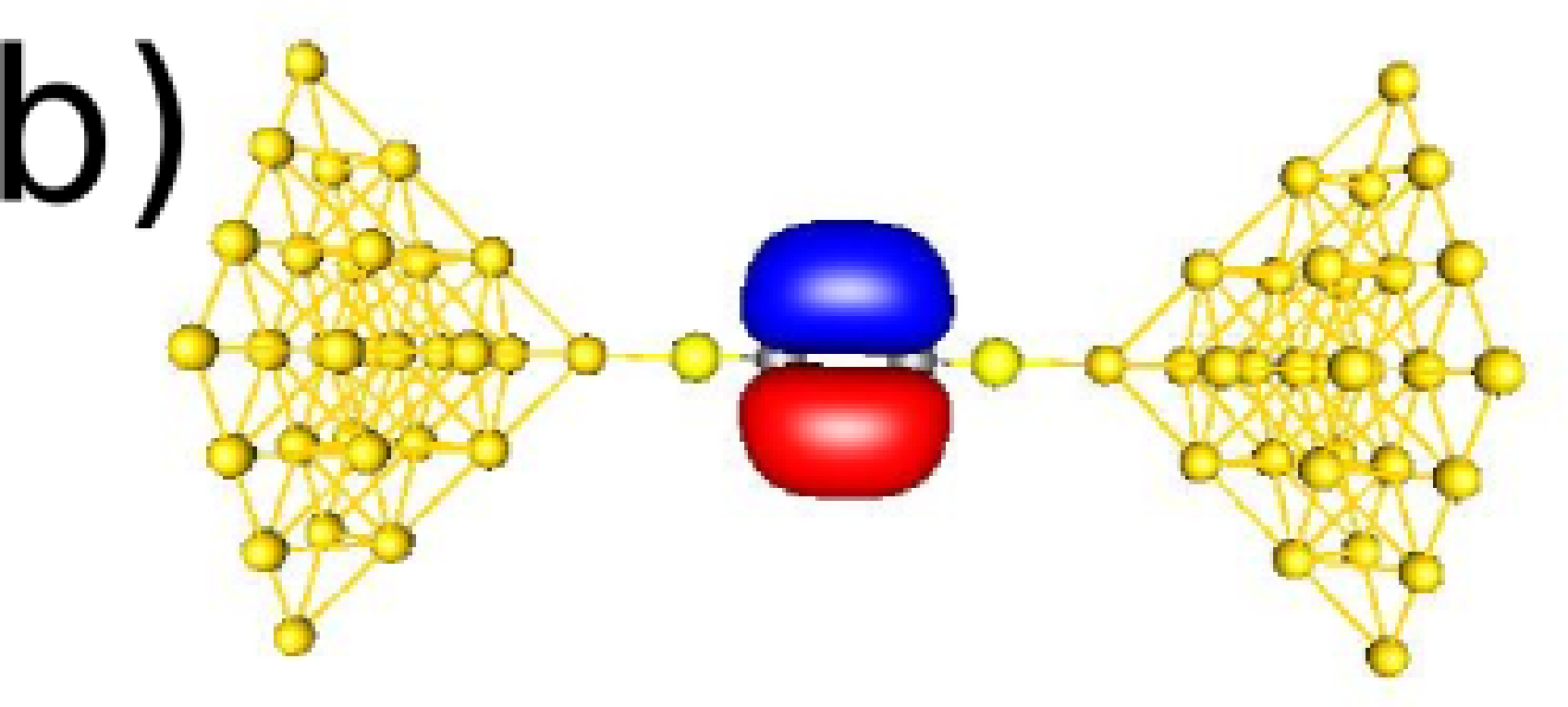}\\
  \includegraphics[height=2.5cm]{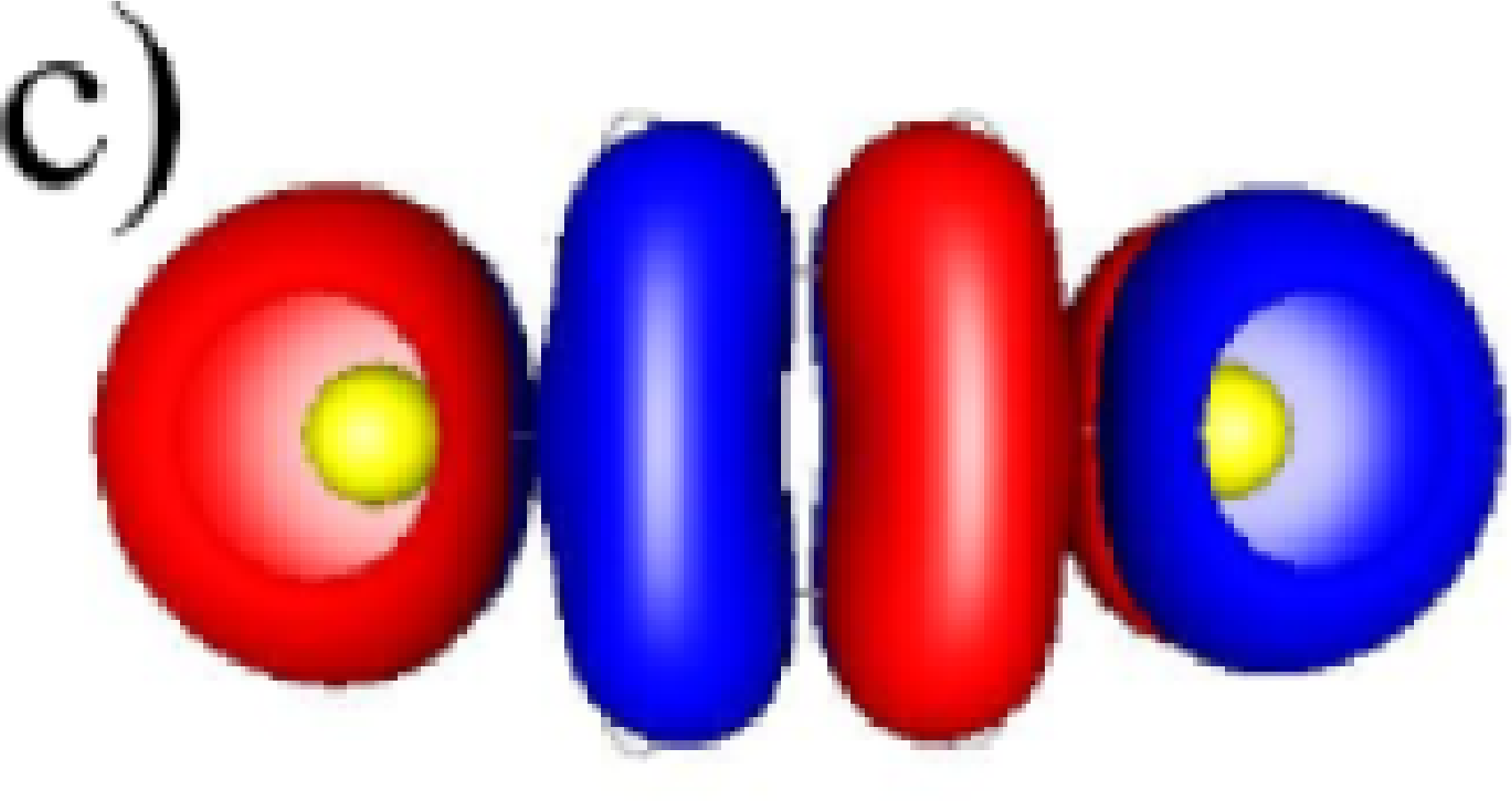}&
  \includegraphics[height=2.5cm]{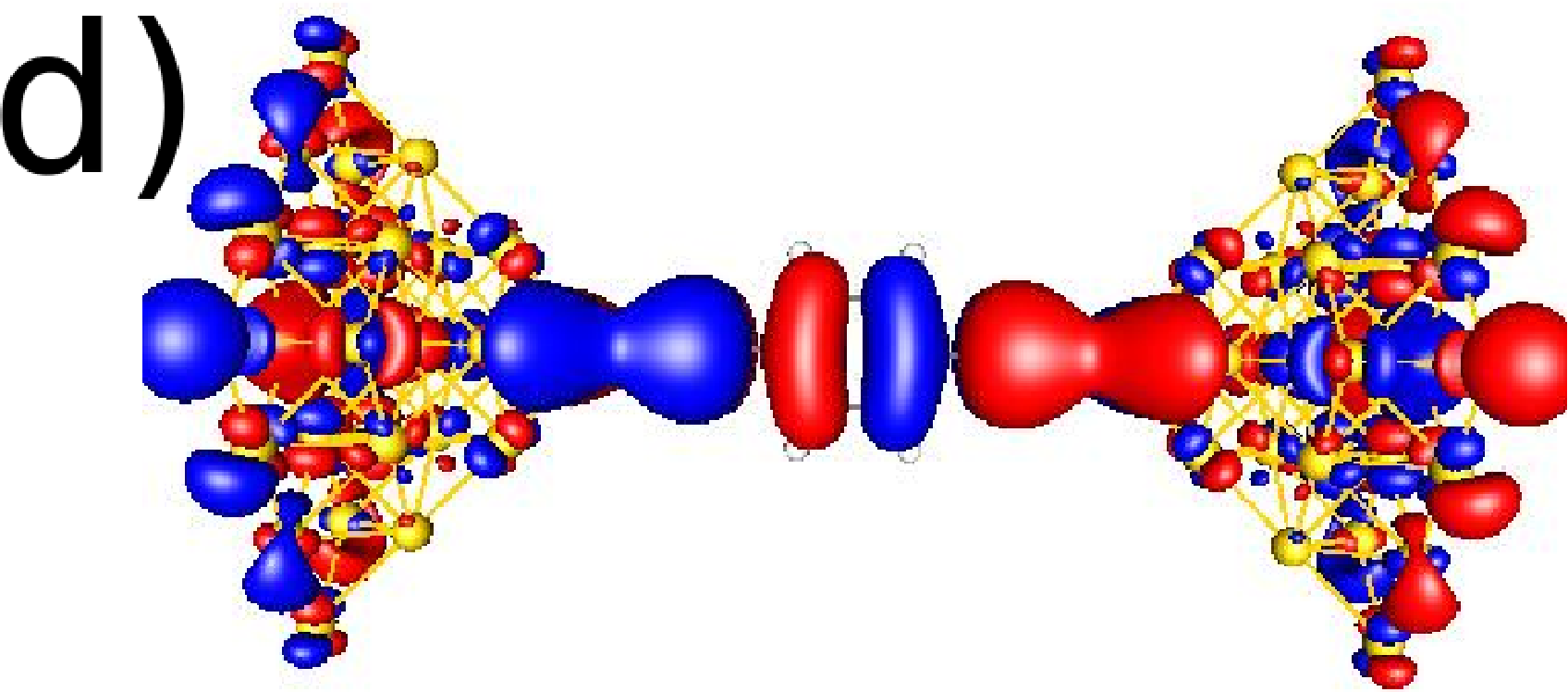}
\end{tabular}
  \caption{Plots of two orbitals, $\psi$,  for di-thiophenyl without
    electrodes (left) and with $N_\mathpzc{E}{=}30$ Au-atoms attached (right).
    Orbital a) hybridizes only very little with the electrodes and is
    almost unaffected after coupling, b). By contrast, orbital c)
    becomes completely delocalized in the metal, see d). 
    Grey (black) colors indicate regions with $\psi$ positive (negative).}
  \label{f1orb}
\end{figure}

Fig. \ref{f1b} shows, how the six orbitals of the molecule in the
gas-phase shift and hybridize with $N_{\mathpzc{E}}$ increasing from
$0,1,5,14,30,71$. For illustration, we have also given
the Kohn-Sham wavefunctions of two representative states in
Fig. \ref{f1orb}.

The overall plot clearly shows, that the original orbitals
survive the coupling to the electrodes and therefore contribute as
resonances to the transport characteristics. 
The initial evolution at small $N_{\mathpzc{E}}$
is not very smooth, which is because a) attaching the first
few Au-atoms cannot be considered a very
small perturbation to the molecular system and b) 
at ``magical'' numbers of atoms, e.g. $N_{\mathpzc{E}}{=}5$,
the electrode configuration is particularly stable. 
These ``stability'' islands are interesting in themselves
but for our present purpose they deliver 
parasitic side effects, since they make it more difficult to
extrapolate the overall flow. When keeping away from
exceptional numbers, e.~g. taking
$N_{\mathpzc{E}}{=}14,30,71$, the evolution shows
the expected smooth behavior. 

We have already mentioned, that the smooth evolution of single
particle levels can also be perturbed, if a prominent molecular level
happens to cross the Fermi energy. This can happen in
a situation where the HOMO is relatively close to $E_{\rm F}$.
Then small fluctuations of the charge
distribution, that occur due to the gradual
appearence of ``evanescent'' modes, i.e.
invading electrode states with
energies in between the prominent molecular orbitals,
can lift the (designated) HOMO above $E_{\rm F}$ 
at certain electrode configurations. 

\begin{figure}[tbp]
  \centering
  \includegraphics[width=1.0\linewidth]{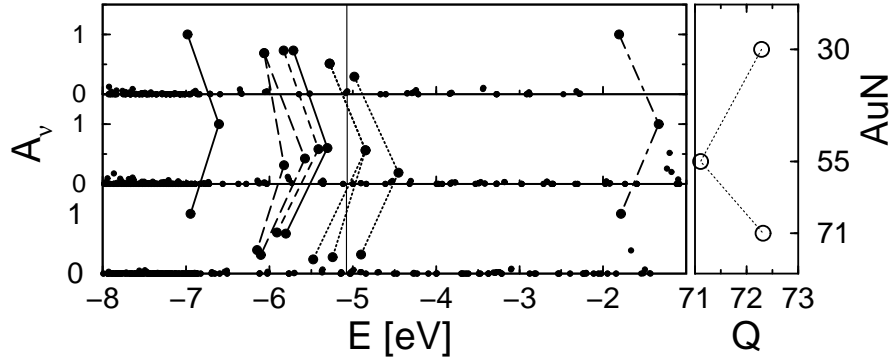}
  \caption{\label{f1c} Plot similar to previous Fig. \ref{f1b} with
    $N_{\mathpzc{E}}{=}55$ included. A right shift of the prominent
    orbitals is caused in the center panel because of
    additional, ``evanescent'' modes appearing. 
}
\end{figure}

This is what is happening with Au55 electrodes,
as can be seen from Fig. \ref{f1c}.
In this case a molecular state
(that turns out to be  localized predominantly on the
S-atoms) 
peaks above $E_{\rm F}$ and therefore is evacuated, leading to a
decrease of the charge accumulated
on the molecule by 1.2$e$.
The fate of this prominent mode that has been expelled from
the region of occupied energy levels
is a fast decay with further increasing $N_{\mathpzc{E}}$,
because of its relatively strong coupling to
the ``invaders'', see Fig. \ref{f1c}.

So far we have witnessed the transformation (or in some cases the
decay) of the states, that occurs as a consequence of
the hybridization of electrode and molecular orbitals.
In a sense, this is the analogue of Fermi-liquid theory.
The Kondo-effect, which in principle could appear for molecular
systems that carry a spin, cannot be understood within
this picture. This is because
the Abrikosov-Suhl resonance is not a shifted molecular
level. Instead, it is a {\em collective phenomenon} and
generated by a large number of electrode
states. Their energies reside inside HOMO-LUMO gap of the ``dressed'' molecule
close to the Fermi-energy. This effect can be seen
with DFT in principle, but the current approximations for the
exchange correlation functional are too crude in order to capture it.
A study with the exact, at present unknown functional
should show satellites at $E_{\rm F}$ produced by
lead states that merge with one another if the system size becomes
large. If sufficiently many of them superimpose,
a sharp peak, the Abrikosov-Suhl resonance,
grows right within the HOMO-LUMO gap.
A further typical characteristics of this collective effect is, that the
resonance position is always the Fermi energy, irrespective of shifts
in the molecular orbital energies, that might be induced by a gate,
for instance.

\subsection{\label{s:IIB} Extended Molecule and $\Sigma_{e\mathpzc{M}}$}

The self energy of the original molecule, $\Sigma_{\mathpzc{M}}$, can
contain a wealth of nontrivial information, it is
not a quantity easy to calculate. This was the message of the
preceeding section. However, 
the situation greatly simplifies, after the molecule has been
redefined. 
Let us consider an extended molecule, $e\mathpzc{M}$,
that comprises in addition
to the original molecule also a ``contact region'', i.~e.
a number $N_{\mathpzc{E}}$ of electrode atoms. 
The Green's function for an extended system, $G_{e\mathpzc{M}}$,
is related to the full Green's function $G$ via a new self energy 
\be
  G^{-1}(E) = G^{-1}_{e\mathpzc{M}}(E) - \Sigma_{e\mathpzc{M}}(E).
  \label{e3}
  \ee
We give two reasons, why it is that $\Sigma_{e\mathpzc{M}}$ is much
easier to handle than $\Sigma_{\mathpzc{M}}$. 

Imagine the extreme limit, in which far more electrons are sitting
on the metal than on the molecule. Then the HOMO of the
big system, $\epsilon_{\rm H}$, 
is given by the Fermi energy of the metal, $E_{\rm F}$,
up to a small uncertainty, which is of the order of the
HOMO-LUMO gap of $e\mathpzc{M}$, $\delta_{\rm HL}$.
In a metal, this gap is inversely proportional
to the number of metal electrons in the calculation,
$\delta_{\rm HL}\propto 1/N_{e\mathpzc{M}}$,
so that the uncertainty of the position of the
Fermi energy with respect to $\epsilon_{\rm H,L}$ 
can be made arbitrarily small.
This is a very obvious advantage. 

More importantly, the flow of the typical
level broadening, $\gamma_{\rm H,L}$
(related to $\Sigma_{e\mathpzc{M}}$),
that is driven by increasing  the number of electrode atoms, $N_{\mathpzc{E}}$,
in the contact region, will lead  us into a very tractable regime as
we shall see now. It is only a fraction $N_{\mathpzc{S}}$ of the $N_\mathpzc{E}$
electrode atoms that is actually connected to the rest of the leads,
the ``outside world''.  Assume, that  $N_{\mathpzc{E}}$ grows in such a fashion, that the 
number of these ``surface atoms'', $N_{\mathpzc{S}}$, does not change,
so we build a quasi-one-dimensional wire.
Then, increasing $N_{\mathpzc{E}}$ implies,
that the fraction of the wavefunction amplitude of extended orbitals
sitting near the surface
decays like $N_{\mathpzc{S}}/N_{\mathpzc{E}}$,
so that $\gamma_{\rm H,L}$ scales like $\delta_{\rm HL}$. 
In good metals the ratio of both energies is of the order of the
metallic conductance
\be
\gamma_{\rm H,L}/\delta_{\rm HL}\sim g\apprge 1. 
\label{e4}
\ee
So, the level broadening of HOMO and LUMO of the extended molecule
always exceeds their separation if the electrodes are made from a good
metal. This is a situation, exactly opposite to the
problematic one, $\gamma_{\rm h,l}/\delta_{\rm hl}\ll 1$, that we have 
encountered before in the context of (\ref{e6b}).
Summarizing, for the extended molecule, $e\mathpzc{M}$,
the following hierarchy of inequalities holds true (\ref{e4a}).
\be
\label{e4a}
           \delta_{\rm HL} \leq \gamma_{\rm H,L} \ll \gamma_{\rm h,l}
           \leq \delta_{\rm hl}.
\ee

The separation of energy scales implied by (\ref{e4a})
is the prerequisite for the real gain
that one makes when one turns to the extended molecule. 
The point is that the fine structure in the
spectrum of $G_{e\mathpzc{M}}$ is on the scale of $\delta_{\rm HL}$.
The anti-hermitian constituent of $\Sigma_{e\mathpzc{M}}$, $\gamma_{\rm H,L}$,
provides the smearing of this fine structure necessary in order to
obtain smooth curves, e.g. for spectral and transmission functions. 
The details of this smearing  have very little impact on the
resulting curves, because the interesting structures
live on energy scales  $\delta_{\rm h,l}$ and $\gamma_{\rm h,l}$,
which exceed $\delta_{\rm HL}$ and $\gamma_{\rm H,L}$ by a
parametrically large factor.

\subsection{\label{s:IIC}$\Sigma_{e\mathpzc{M}}$
  and Absorbing Boundary Conditions}
  
There is yet another, perhaps particularly intuitive way to understand
the principal difference between $\Sigma_{\mathpzc{M}}$ and
$\Sigma_{e\mathpzc{M}}$. It will serve as a motivation for the
proposed approximation (\ref{e6}).

Let us assume, we opted for an investigation of
transport properties in the time domain, e.~g. by
propagating wavepackets. Then, we 
would study the time evolution of a wavepacket,
that is localized at $t{=}0$ at some initial position
on the molecule. In particular, we can investigate
how wavepackets leak out of
the molecule into the contacts such that they gradually disappear.
When performing such an investigation systematically
in the presense of leads, one can in principle
collect enough information in order to reconstruct
the Fourier transform of the retarded Green's function,
$G(t)$. 

There is a condition on the observation time, $T$.
In order to have  an energy resolution $\gamma_{l,r}$
we need $T \apprge \gamma_{l,r}^{-1}$.
This  comes for us with trouble,
if the contact size maintained in our calculation
is not sufficiently large. After some time,
the leakage will hit the outside walls of the contact, reflect back
off them and finally, after the ``dwell'' time, $\tau_D$, it will rearrive
at the molecule; we calculate $G_{e\mathpzc{M}}(t)$
instead of $G(t)$. The energy resolution, that we achieve with such a 
calculation, can not exceed $h/\tau_{\rm D}$. 
The best that we can hope for is $\tau_{\rm D}\sim \delta_{\rm HL}^{-1}$
but only in a case, where the contact acts as a fully chaotic cavity.
At longer times the signal from the decaying wavepacket
will be superimposed by the cavity modes that describe how the
wavepacket sloshes back and forth inside the electrodes.
We shall demonstrate this effect
in Sec. \ref{s:III} looking at an explicit example.

The salient point we wish to make is,
that the minimum dwell time in the cavity should be so long that the
wavepacket has enough time to evacuate before the molecule is being
refilled again from the backscattered modes:
\be
\label{e6e}
\tau_D\gamma_\mathpzc{l,r}\gg 1.
\ee
There is a very elegant and powerful procedure that  
eliminates spurious cavity modes so that 
the condition (\ref{e6e}) is always satisfied:
one introduces {\em absorbing boundary conditions} (abc)
in some regions of the cavity. These
``surface'' regions should 
swallow incoming signals, i.e. wave packet amplitude, 
and thus properly mimic escape to infinity. If
the wavepackets, bouncing hence and forth inside the cavity, are
completely eliminated before the return
time $\tau_{\rm D}$ has passed, no trace of the finite
cavity/electrode size will be left and $G(t,{\bf x},{\bf x'})$
can be reconstructed. This is exactly how adding the self energy
$\Sigma_{e\mathpzc{M}}$ works and 
nothing more than this is implied, if the contacts are sufficiently
large. Therefore, we are entirely free to replace the
exact boundary conditions $\Sigma_{e\mathpzc{M}}$ by
any other ones, provided they absorb sufficiently fast
(and do not disturb the immediate vicinity of the
molecule-electrode junction).

This is the idea underlying the step proposed in (\ref{e6}).
From what has been said above, it should have become clear that
this ansatz is actually not just a good  approximation,
but it will give exact results, if the
number of metal atoms $N_{\mathpzc{E}}$ is sufficiently large
and the damping function has been chosen well enough.
The examples given in Sec. \ref{s:IV} suggest, that
a relatively small number of contact atoms $\sim 10-20$
can already give reasonable results. 


\section{\label{s:III} Toy Models}

In this section we are going to analyze two toy problems as test
cases, 
namely the conductance of an $L$-site tight binding wire, clean and 
in the presence of an obstacle, that mimics a molecule.

We will begin with a single channel wire and show, that
the technique introduced with (\ref{e6b})
delivers the correct answer.
This test case is interesting, because
 a) the numerical results can be compared to analytical formulas
 and b) it is particularly difficult,
in the sense that the dwell time, $\tau_D\approx L/2v_{\rm F}$, is
(untypically) small ($v_{\rm F}$: Fermi velocity).
To illustrate the method further, we also apply it to
a wire with four channels and show, that not only the
conductance but also more complicated quantities, like 
the local current density, can be obtained. 

\subsection{Single Channel Tight Binding Wire}
\begin{figure}[htb]
  \centering
  \includegraphics[width=0.9\linewidth]{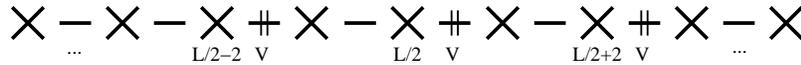}
  \caption{Single channel tight binding wire with triple barrier
    realized by weak links (indicated by vertical double lines) as used in model
 calculations. }
  \label{sctbw}
\end{figure}

\subsubsection{Models and Analytical Results}

\begin{figure}[t]
\includegraphics[angle=270,width=0.9\linewidth]{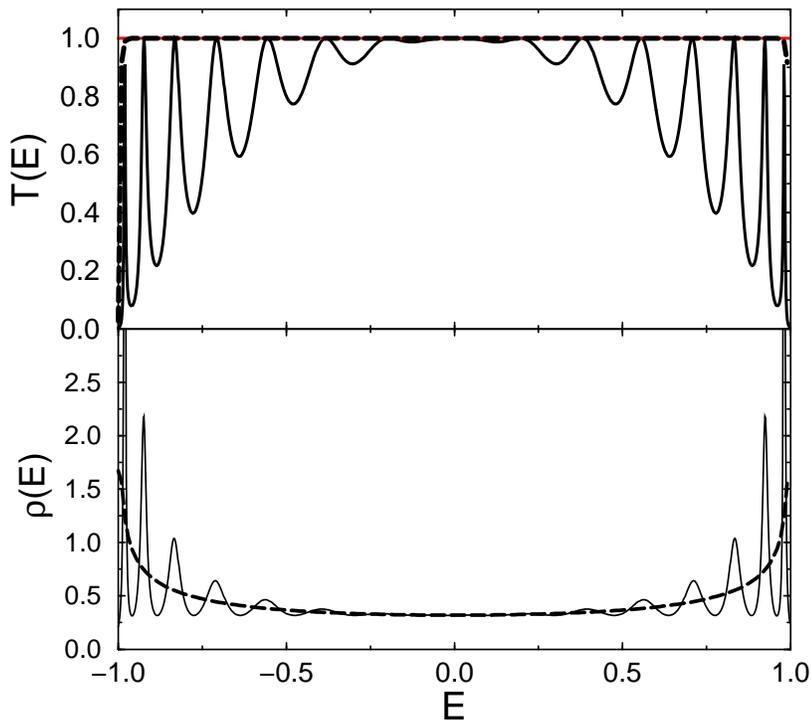}%
\caption{\label{f1} Clean single channel
  tight binding wire, upper panel:
  transmission. Analytical result (\ref{e9a}) (solid,thin)
  and numerical result for $L{=}256$ with fully absorbing
  boundary conditions (dashed; $i_\mathpzc{S}{=}32$,
  $\beta{=}0.3$, $ \eta{=}1$). Also shown is numerical result with partially absorbing
  boundary conditions for $L{=}16$ (solid, wiggly line, $i_\mathpzc{S}{=}1$, $
  \beta{=}\infty$, $ \eta{=}1$). Lower panel: density of states
  corresponding to the transmission curves shown in upper panel.
  }
\end{figure}

The model Hamiltonian of the clean tight binding wire is given by
\be
H = -t/2 \sum_{i{=}1}^{L} c^\dagger_{i} c_{i+1} + c^{\dagger}_{i+1}
c_{i}
\label{e7}
\ee
for spinless, non-interacting electrons. The corresponding dispersion relation reads
\be
\epsilon_{k}=-t \cos(ka),
\ee
where $a$ denotes the lattice constant, and for the density of states
one has 
\be
    \varrho(E) = \frac{1}{\pi t \sqrt{1-(E/t)^2}}.
\label{e9}
\ee
It exhibits the usual van Hove singularities at the edges of the band,
which can also be seen in Fig. \ref{f1}.

In order to calculate the transport coefficient of the $L-$chain
defined in (\ref{e7}) we should couple it adiabatically to a left and a right
hand side reservoir. This can be done by attaching further
half-infinite tight binding chains to the right and the left \
of the $L-$chain. The combined system is a perfect 1$d$ crystal. 
Its Bloch waves travel without any backscattering through the
$L-$chain and therefore it is a perfect conductor with transmission
unity:
\begin{equation}
  \label{e9a}
T(E){=}1 \qquad |E|<t
\end{equation}

Next, let us insert an obstacle, e.~g. a strong triple barrier
(see Fig.~\ref{sctbw}),
into the wire, a situation that still can be understood in all detail.
The corresponding Hamiltonian $H_{e\mathpzc{M}}$
is realized with  hopping
amplitudes occurring in (\ref{e7}) that
take the values
$t{=}0.05$ at the pairs of sites
$(i_{\pm}{=}L/2, i_{\pm}{=}L/2\pm 2)$
and $t{=}1$ everywhere else.

\begin{figure}[htb]
\includegraphics[width=1.0\linewidth]{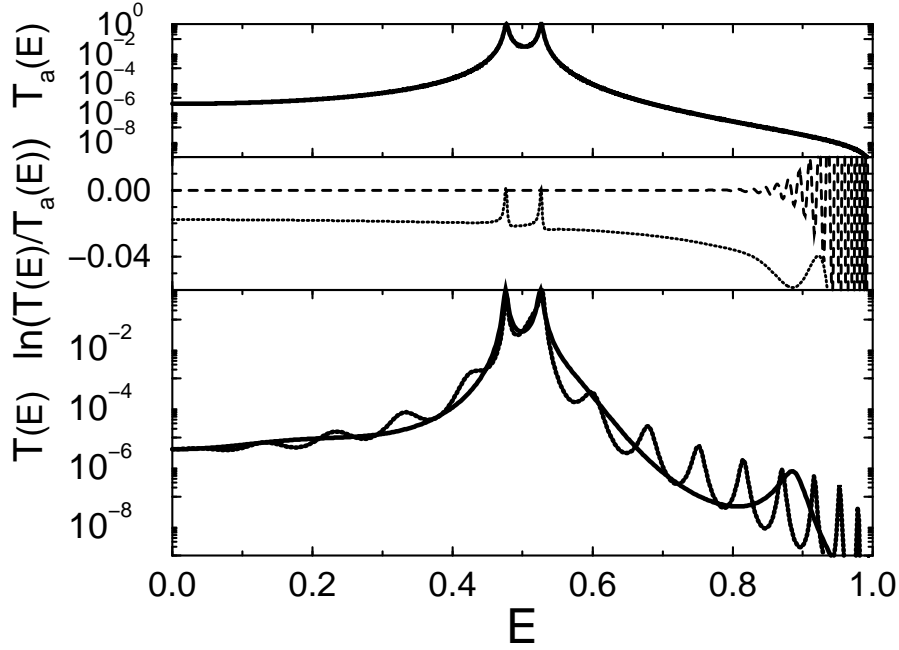}%
\caption{\label{f2} Transmission of a single channel wire
  with a strong triple barrier (see text). Upper panel:
  exact analytical result. Center: deviation of numerical results from 
  analytical calculation: $\ln\frac{T(E)}{T_a(E)}$
  ($L{=}256$, $i_\mathpzc{S}{=}32$, $\beta{=}0.3$, $\eta{=}1$, dashed;
  $L{=}64$, $i_\mathpzc{S}{=}16$, $\beta{=}0.3$, $\eta{=}1$, dotted)
  with absorbing boundary conditions.
  Lower panel: partially absorbing boundary conditions
  (dotted, wiggly line: $L{=}16$, $ i_\mathpzc{S}{=}1$, $ \beta{=}\infty$, $
  \eta{=}1$)
  (solid, smooth line: $L{=}64$, $ i_\mathpzc{S}{=}1$, $ \beta{=}\infty$, $ \eta{=}1$)}
\end{figure}

The triple barrier has two eigenstates,
a symmetric and an anti-symmetric one, which are energetically
nearly degenerate since the center barrier is high.
The energy is given approximately by the width of the
double well inside the outer barriers, $3a$.
It corresponds to a wavenumber
$k=\pi/3a$ which in turn implies a resonance
energy close to $t/2$.
Therefore, the transmission characteristics of the
triple barrier should exhibit a superposition of two Lorentzians,
one slightly below and one slightly above $E=t/2$. 
These are the features that can indeed be seen in the
analytical result for the conductance
(valid in the limit of weak coupling, $V\ll t$) 
  \be
  T(E) = \sum_{\alpha{=}\pm} \frac{1}{1+\gamma^{-2}(E-E_{\alpha})^2}
\ee
where $E_{\pm}{=}\cos(\pi/3{\pm}V\sin(\pi/3)/3)$, $t{=}1$ and
$\gamma{=}V^2\sin(\pi/3)^3/24$. A derivation can be found in 
appendix \ref{a1}. Here, we have reproduced for better clarity
only an expansion of the exact answer, given in (\ref{ea7}),
which is valid in the vicinity of the resonances. 
We also display the exact transmission in Fig. \ref{f2}, upper panel.

\subsubsection{\label{s:IIIA2}Green's Function Method with Absorbing Boundary Conditions}

The transmission of the combined system
-- wire plus triple barrier -- has been given already
in (\ref{ea}). In the present case it reads 
\be
g(E) = \text{Tr}_\mathpzc{eM}\  G \Gamma_{\mathpzc{L}} G^\dagger \Gamma_{\mathpzc{R}},
\ee
where the definition of $G$, (\ref{e1c}), implies:
\be
     G^{-1} = E - H_{e\mathpzc{M}} - \Sigma_{\mathpzc{L}} - \Sigma_{\mathpzc{R}}.
\label{e11}
\ee
The Hamiltonian $H_{e\mathpzc{M}}$ of the extended molecule
is given with Eq. (\ref{e7}) and the trace $\text{Tr}_{e\mathpzc{M}}$
is over the corresponding Hilbert space.
The operators $\Gamma_{\mathpzc{L,R}}$ are related to those pieces,
$\Sigma_{\mathpzc{L,R}}$, of the self energy, $\Sigma_{e\mathpzc{M}}$,
which describe the level broadening due to the coupling
to the left ($\mathpzc{L}$) and right ($\mathpzc{R}$) leads:
\be
    \Gamma_{\mathpzc{L,R}} = i(\Sigma_{\mathpzc{L,R}} - \Sigma_{\mathpzc{L,R}}^\dagger).
\label{e12}
\ee
The precise form that the $\Sigma_{\mathpzc{L,R}}$ take, depends on 
how we couple the wire to the external leads. In the spirit of
section Sec. \ref{s:IIC}, we simply define $\Sigma_{\mathpzc{L,R}}$
as follows:
\bea
    \Sigma_{ij; {\mathpzc{L}}} &=& i \eta_{i} \delta_{ij}; \qquad \eta_{i} {=}
    \eta/(1+\exp{\beta(i-i_\mathpzc{S})}) \label{e14} \\
    \Sigma_{ij; {\mathpzc{R}}} &=& i \eta_{i} \delta_{ij}; \qquad \eta_{i} {=}
    \eta/(1{+}\exp{\beta(L {-} i {-} i_\mathpzc{S})}). \label{e15}
\eea
Three parameters have been introduced: $\eta$ is the atomic leakage rate for
all those atoms that are fully coupled to the outside; $i_\mathpzc{S}$
describes the number of surface atoms on either side of the molecule; $\beta$ is the
adiabaticity parameter, that models a smooth transition into the external wire. 

Fig. \ref{f1}, upper panel,
displays the result of this procedure for the transmission
of the clean wire. As expected,
the exact result (\ref{e9a}) is recovered in the
case with perfectly absorbing boundary conditions (abc).
For comparison, we also show a trace corresponding to
incomplete absorption. The cavity modes manifest themselves
in the transmission characteristics as relatively
sharp resonances.
In order to highlight this aspect, the lower panel of
Fig. \ref{f1} also shows the density of states of the wire. 

In Fig. \ref{f2} we present the transmission of the single channel
wire with a triple barrier implantation.
As can be seen from the upper panel, the agreement between the
wire with fully abc and the analytical
result is perfect.
Once more we also display traces that result from a
calculation with imperfectly absorbing boundaries.
Traces for two different cavity sizes, $L{=}16,64$, are given.
Like in the previous case, Fig. \ref{f1}, the cavity eigenmodes
give rise to system size dependencies, which are the
spurious resonances in the transmission characteristics. 
By contrast, no remnant of the system size is
left if perfect abc are used, see upper panel
traces for $L{=}64,256$. 

Let us emphasize, that the good quality of the results
Fig. \ref{f1} and \ref{f2} is not a consequence of
fine-tuning parameters. We have ascertained, that the traces
corresponding to perfectly abc 
are stable against variations at least in the parameter range
$L{=}64{-}512$, $\eta{=}0.1{-}10$,
$\beta{=}0.03{-}0.5$, $i_\mathpzc{S}{=}8{-}64$.

In the test cases presented in this subsection, analytical results
have been available in order to demonstrate that the choice of parameters
associated with the absorbing boundary conditions was appropriate.
In more realistic situations analytical results are almost never
available. Therefore additional criteria have to be given so as to
establish that a certain choice of boundary conditions indeed
provides sufficient absorption. 
The basic rule is, that a good implementation will
yield transmission curves that are (largely)
independent of the size $N_{e\mathpzc{M}}$ of the cavity,
of the choice of surface atoms inside the cavity
and of the atomic leakage rate $\eta$. 
A calculation, that satisfies these requirements,
is (usually) quite reliable.

\subsection{Local Currents in a Many Channel Tight Binding Wire}

As a further application of our method, we calculate the local current
density, $j_{\mu}$,  in a multi-channel wire. We begin by deriving a
general formula relating $j_{\mu}$ to the Green's functions and self
energies, that have been calculated in the preceeding section.
Thereafter, we shall illustrate the result by calculating the
local current distribution within a double well embedded in a four channel
wire. 

\subsubsection{Lattice Current Density in Terms of Green's Functions}

To start with, we consider the general model Hamiltonian
\be
H = \frac{1}{2} \sum_{\nu,\nu'} t_{\nu,\nu'}  \ c^\dagger_\nu
c_{\nu'} .
\ee
The multi-index $\nu$ comprises the longitudinal, $i_{\nu}$, and
transverse, $\ell_{\nu}$, wire coordinates.  
An expression for the local (longitudinal)
current density may be obtained from the
time dependent local density:
\be
    \dot n_{\mu} = \frac{i}{\hbar} [H, c^\dagger_{\mu} c_{\mu}]
                 = -\frac{i}{2\hbar} \sum_{\nu} 
                 t_{\mu\nu} c^\dagger_{\mu} c_{\nu} - t^*_{\mu\nu}c^\dagger_{\nu}c_{\mu} . 
\ee
The component of the local particle current in the longitudinal
direction (right), $j_{\mu}$,
is given by the difference of those hopping events
that enter site $(i_{\mu},\ell_{\mu})$ from the left and leave there again:
\be
     j_\mu = -\frac{i}{2\hbar}
     \sum_{i_{\nu} < i_{\mu}} \sum_{\ell_{\nu}}
     t_{\mu \nu} c^\dagger_{\mu} c_{\nu} -
     t^*_{\mu\nu}c^\dagger_{\nu} c_{\mu} . 
\label{e18}
     \ee
The expectation value of this operator is readily expressed 
in terms of the Green's function \cite{meir92},
$G^<_{\nu\mu}(t,t'){=}i\langle c^\dagger_{\mu}(t')c_\nu(t)\rangle$,
\be
\langle j_\mu \rangle = -\frac{1}{2\hbar} \int \! \frac{dE}{2\pi} \
\sum_{\nu}^{\mathpzc{L}_\mu} t_{\mu\nu} \ G^{<}_{\nu\mu}(E) - t^*_{\mu\nu} \
G^{<}_{\mu\nu}(E). 
\ee
In order to simplify the notation, we have
introduced a name for sums like the one appearing in (\ref{e18}),
which are restricted to the left/right half space: $\sum^{\mathpzc{L,R}}$.  
Since we only consider non-interacting fermions, $G^{<}_{\nu\mu}$ takes a
particularly simple form (e.~g. \cite{datta95})
\be
G^{<} = i G\left( f_{\mathpzc{L}} \Gamma_{\mathpzc{L}} + f_{\mathpzc{R}} \Gamma_{\mathpzc{R}}
\right) G^\dagger,
\ee
which in the limit of small voltages $V$  leads to the following
formula for the local charge current distribution
$j(i_\mu)=\sum_{\ell_\mu} j_\mu$:
\be
e\langle j(i_\mu) \rangle {=} {-}\frac{i e^2}{2h} \!V \!\int\!\! dE \
\frac{df}{dE} \ \sum_{\ell_\mu} \sum_{\nu}^{\mathpzc{L}_\mu} t_{\mu\nu}
\left[ G
\left( \Gamma_{\mathpzc{L}} {-} \Gamma_{\mathpzc{R}} \right)
                  G^\dagger \right]_{\nu\mu}.
\label{e21}
\ee
(When writing this expression, we have assumed
for simplicity that $H$ is time reversal invariant,
so that $t$ is real and $t, G$ and $\Gamma$ are symmetric matrices.
Also, $f_{\mathpzc{L,R}}$ denote the Fermi-Dirac distribution of
quasi-particles in the left and right hand side reservoirs.)

\subsubsection{Application: Double Dot Inside a Four Channel Wire}

\begin{figure}
\centering
\includegraphics[width=0.75\linewidth]{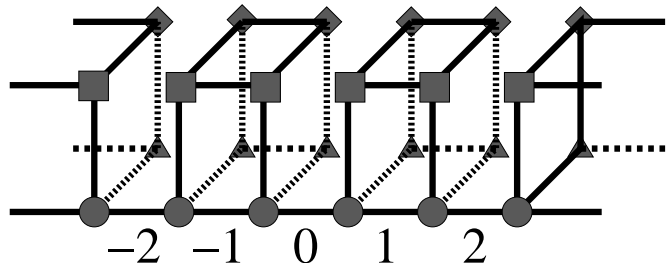}%
\caption{\label{f1a} A double dot inside a tight binding wire consisting of
  four coupled strands (indicated by different symbols).}
\end{figure}

In order to give an example for the usefulness of 
(\ref{e21}), we calculate $j_{\mu}$ for a four channel wire
with a double well barrier. The example illustrates, that the average
current in the wire is a sum of contributions. They can undergo strong
spatial fluctuations which are of the order of the mean current
and in particular they can be positive or negative (backflow).
It is only the sum of all of them which is
independent of the longitudinal spatial coordinate.

The clean 4-wire considered here is made up from four strands,
which are the 1-wires given in (\ref{e7}).
These 1-wires are arranged in a 4-fold cylindrical geometry and
in transverse direction only nearest neighbors are being coupled,
see Fig. \ref{f1a}.
The coupling between all nearest neighbor pairs
has been chosen $t{=}2$, except for those 9 pairs,
that form the double well. They have $t{=}0$.
The position of these ``defects'' can be given in terms of the
longitudinal site index, $i$, and the transverse index $\ell$, that
labels the constituting 1-wires in a clockwise fashion:
$\ell=0,1,2,3$.
We have switched off three couplings at and near the center ($i_C{=}L/2$)
of the wire, $i_C,i_C\pm2$,  
with site indices $\ell{=}1,2,3$.  
After the definition of the model Hamiltonian, $H_{e\mathpzc{M}}$,
we also give the left
and right contributions to the self energy $\Sigma_{e\mathpzc{M}}$, 
\be
     \Sigma^{\ell\ell'}_{ij;{\rm X}} = \delta_{\ell\ell'}
     \Sigma_{ij;{\rm X}}, \qquad {\rm X}={\mathpzc{L,R}}
\ee
with a self energy per strand, $\Sigma_{ij;{\rm X}}$,
as in (\ref{e14}, \ref{e15}).

\begin{figure}[t]
\includegraphics[width=\linewidth]{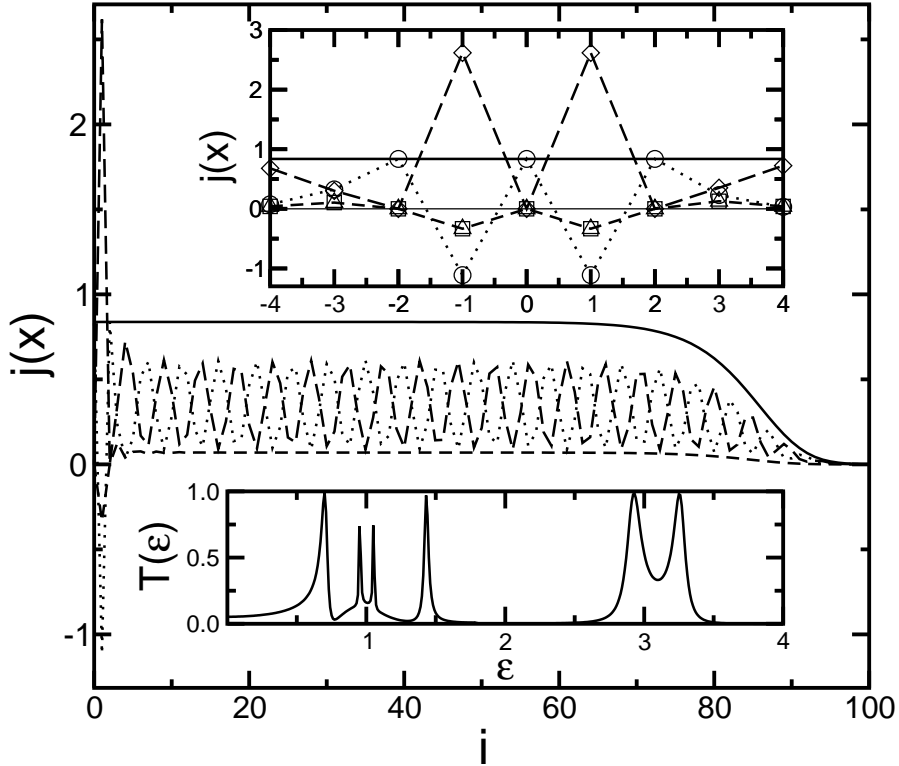}%
\caption{\label{f4} Local current density 
  inside a four channel wire
  ($L{=}256$, $i_\mathpzc{S}{=}32$, $\beta{=}0.1$, $\eta{=}1$)
  with a double dot (Fig. \ref{f1a}) at $\epsilon{=}0.7$
  very close to a resonance of the transmission (see lower inset).
  Upper inset: behavior near the dots (position $\pm2,0$).
  Channels $1,3$ ($\Box,\triangle$,
  dashed line; same current density)
  and $2$ ($\diamond$,long dashed) are blocked,
  channel zero $0$ ($\circ$, dotted line) remains open.
  Inside the wells, $j_2$ overshoots the transmission (solid line)
  by 300 \%,
  so that local backflow in the other current channels is generated.  
  Main: sum of local currents (solid line) is conserved,
  i.e. independent of position (except for the surface region, where
  by adding the self energy leakage has been introduced).
  }
\end{figure}

Again, we may employ (\ref{e11}, \ref{e12}), so that
the numerical evaluation of (\ref{e21})
is straightforward ($T{\to}0$). 
Fig. \ref{f4} shows the induced current density per applied voltage
near the resonance energy, $\epsilon=0.7$.
Before we discuss this result in more detail,
we first address the structure of the transmission function which is
also displayed in Fig. \ref{f4}, lower inset.
It consists of three pairs of resonances, each pair resembling the
symmetric and anti-symmetric eigenstate of the (isolated)
double well. The pair of peaks closest to the band edges
originates from hybridization of these states
with those wire modes, that have a
wavenumber matching $\pi/2a$ and an $s$-wave type symmetry in
transverse direction. In these two peaks the current density is
homogeneously distributed among the four constituting wires.
The last sentence is not true for the remaining two pairs of
resonances with $p$-wave character,
where the current flows mainly in one of the wire pairs, either
(0,2) or (1,3). Evidently, the two resonances closest to $\epsilon{=}1$
belong to the second category, since these are much sharper and less
well split than all the others.

Now, we can come back to the strong oscillations seen in the local
current density
of the resonance closest to the band center, main panel of
Fig. \ref{f4}. There, the current flow is mostly
in the (0,2) pairs.  Since due to the barrier these two channels are 
not symmetry equivalent, the current densities $j_{i0}$ and $j_{i2}$ can pick up
different dependencies on the longitudinal wire coordinate, $i$.
In fact, this must be the case since at the barrier position 
$j_0{=}1$ while $j_2{=}0$. 

The phase locking between the local currents flowing along the
(0,2) pair of strands has
an interesting effect on the current flow inside each
well: in this region,
the component $j_2$ acquires a value three times exceeding the average
current flow. This value is compensated by a backflow in the other
channels, so that a current vortex develops.

\section{\label{s:IV} Test Cases from Quantum Chemistry Calculations}

The purpose served by the tight binding calculations of the previous
chapter was to demonstrate the principle. High precision in
the calculations performed there was relatively easy to achieve,
because the parameter space representing a perfect 
separation of energy (or time) scales was
well accessible to numerical methods. 

In the case of quantum chemistry calculations
that are feasible at present,
the accessible system sizes often are
not large enough in order to achieve
such a clear scale separation.
What we demonstrate in this section is,
that our method operates reasonably well, also in a practical
situation, where scale separation is not perfect.

To this end, we shall consider two extreme cases, a short gold
chain with a conductance $g\sim 1$ and a molecule, di-thiophenyl with
$g\ll 1$. Both objects are coupled to the tip of
two tetragonal bipyramids of 14 Au-atoms each,
that represent the extension modelling a piece of the
electrodes, see Fig. \ref{f5a}.
We define an effective single particle Hamiltonian, $H_{e\mathpzc{M}}$,
for these systems in the way explained in Sec. \ref{s:IIA4}. 

The proposed construction mechanism has been formulated in
coordinate (real) space. Therefore it matches particularly well with
quantum chemistry calculations, that are using the local atomic
basis sets $|X,\ell\rangle\equiv |b\rangle$ also introduced in Sec. \ref{s:IIA4}. 
Adopting (\ref{e6}) to the present case, 
we introduce  the following self energy:
\be
    \Sigma_{\rm X} = i \eta \sum_{b,b'}^{\mathpzc{S}_{\rm X}} |b \rangle \
          [S^{-2}]_{bb'} \ \langle b'|, \qquad {\rm X}={\mathpzc{L,R}} .
          \label{e24}
\ee
The overlap matrix, $S_{bb'}=\langle b|b'\rangle$, that appears
in this expression takes care of the fact, that
basis states belonging to different atomic
sites will not be orthogonal in general. 
$\Sigma_{\mathpzc{L,R}}$  is local, i.e. diagonal in the atomic basis set
$|b\rangle$, in full analogy to
(\ref{e14}, \ref{e15}). Again the important input
is in the strength and spatial modulation of the leakage function.
In the present case, we choose it to be a constant, $\eta$,
for a subset $\mathpzc{S}_{\mathpzc{L,R}}$
of ``surface atoms'' and zero for all the others. 
In our calculations we take these sets to be the two layers of the
pyramid ($3\times 3$ and $2\times 2$) that are farthest from the
molecule, see Fig. \ref{f5a}.

\begin{figure}
\includegraphics[width=\linewidth]{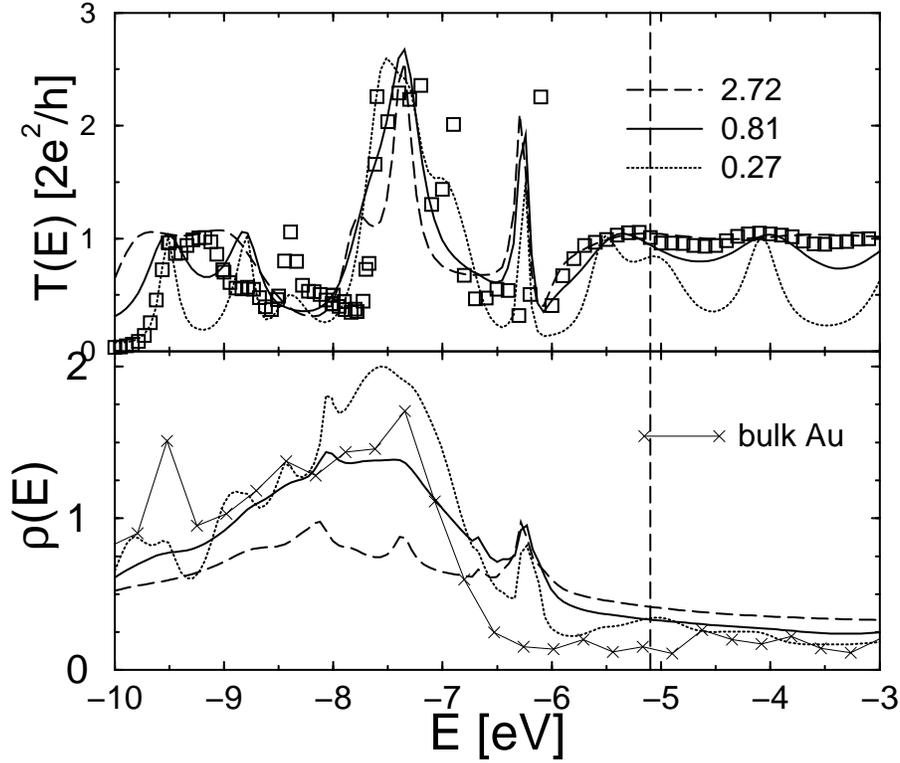}%
\caption{\label{f5} Upper panel: Transmission of
  a two atom Au-chain linked to the tip of two
  tetragonal pyramids of 14 Au-atoms each.
  Self energy (\ref{e24}) has been used. Three traces with
  different $\eta$ are shown: 2.72 eV (dashed), 0.81 eV (solid), 0.27 eV
  (dotted). The $\Box$-symbol indicate the result with the canonical coupling and
  much larger pyramids with 55 atoms \cite{evers03prb}. 
  Lower panel: density of states calculated with the Green's
  functions used in the upper plot. Also the density of states for
  bulk gold is shown (symbol $\times$). 
  }
\end{figure}

\subsection{\label{ss:IIIA} Transmission of Au-Chain}

We begin our analysis with the two atom Au-chain.
The transmission of such chains has been studied
by various groups before \cite{brandbyge02,palacios02}.
It is well known that reproducing the correct transmission curve is a
sensitive test on the quality of the damping $\Sigma_{e\mathpzc{M}}$
of the model. 
In Fig. \ref{f5}, upper panel, we depict the transmission as obtained
from  the model (\ref{e24}). For comparison,
also plotted is a result obtained from a
much larger system with a self energy calculated directly from
(\ref{e1b}) (where $G_\mathpzc{S}$ has been replaced by the
bulk Green's function $G_\mathpzc{B}$) \cite{evers03prb}.
Agreement for the two larger values of the coupling amongst each
other and also with the original curve is  established reasonably
well with deviations typically less than 10 \%. 

As was to be expected, the window of $\eta$-values
in which one finds good quantitative agreement between the various
curves is relatively small. This is simply because of the very 
small cavity size. In fact, the ``molecule'' defined by the
two Au-atoms in line, see Fig. \ref{f5a}, 
are separated from  the effective surface regions
$\mathpzc{S}_\mathrm{L,R}$ by only one gold atom.

In the surface regions, the model self energy somewhat
modifies the local material parameters,
like the density of states (DoS).
This can be seen in Fig. \ref{f5}, lower panel. The total DoS
is strongly dominated by surface atoms. It 
has a dependency on $\eta$ that is much stronger than the one of
the transmission Fig. \ref{f5}, upper panel.
Note, that this modification 
will have a substantial impact, if one
were to set up a self consistent calculation
with a local density obtained from the
dressed Green's function (\ref{e1c}).
The present setup is not
(and in fact does not need to be)
self consistent in this sense, and therefore
the modification of the surface spectral function
will be without consequences for
the transport calculations proposed here. 

We comment on the absence of a adiabaticity parameter $\beta$ in
the definition of the self energy (\ref{e24}).
The pyramids simulating the electrodes 
act as resonating cavities in the same way that the single channel tight
binding wire does, c.f. Sec. \ref{s:IIIA2}. However, the tight binding
wire was special, in the sense that the number of surface atoms
coupling to the infinite tight binding chain 
was only two, independent of the volume of the wire, $L$. For this reason
the resonator modes had to be eliminated by introducing the
adiabaticity parameter $\beta$. In higher dimensions, the ratio of
contact surface to volume is much more favorable. Therefore the surface
damping of the resonator modes is much stronger --
no real need to introduce a $\beta$-parameter here.

\subsection{\label{ss:IIIB} Transmission of Di-Thiophenyl}

Finally, we apply our construction for the self energy to the paradigm
of computational molecular electronics, the di-thiophenyl
system. Again, the electrodes are modelled by the same pair of
14-Au pyramids, that we have used before for the 2-Au-chain.
Accordingly, the construction of the model Hamiltonian and,
in particular, the self energy are just as in the previous
section.

We will investigate two slightly different situations, where 
the sulphur atom, that ties the benzene to the Au-contact surface,
once connects to a single Au-atom and once to three of them.

\subsubsection{\label{sss:IIIB1}${\rm S-Au}_1$ Coupling}

\begin{figure}
\includegraphics[width=1.0\linewidth]{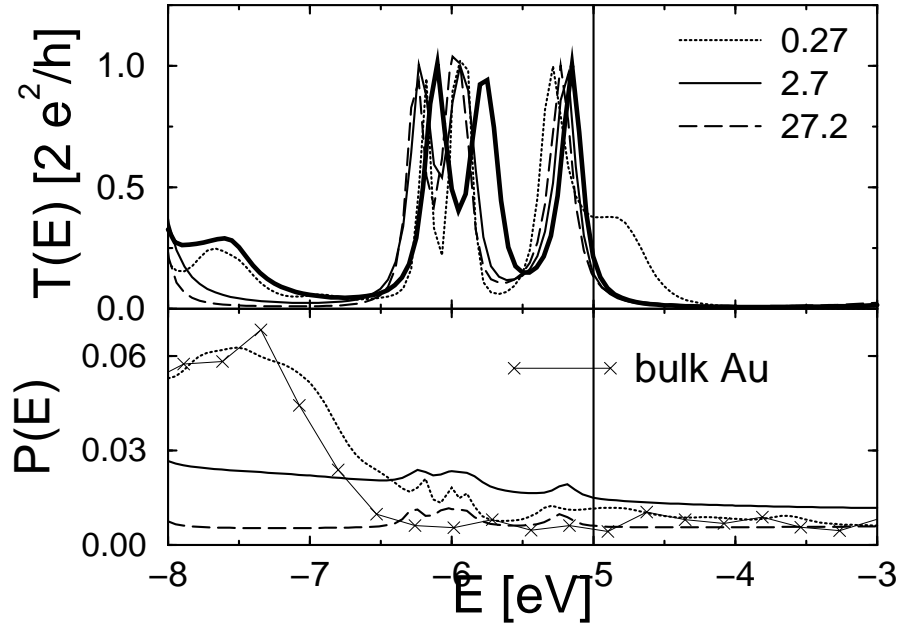}%
\caption{\label{f6}  Upper panel: transmission of di-thiophenyl
  with a ${\rm S-Au}_1$ coupling  (see Fig. \ref{f5a}).
  for different damping parameters $\eta$: 0.27 eV (dotted), 2.7 eV (solid)
  and 27.2 eV (dashed). Also shown 30-Au
  pyramids with conventional coupling (\ref{e24}).
  Lower panel: probability density $P(\epsilon)$ for eigenvalues
  (proportional to the DoS). Three traces correspond to the
  transmission lines of upper panel. Symbols $\times$ indicate
  $P(\epsilon)$ of bulk gold.}
\end{figure}

The atomistic setup, that we consider in this subsection, is
presented in Fig. \ref{f5a}. The S-atom acts as the barrier that
disconnects the conjugated $\pi$-system of the phenyl ring from the
Au-atom forming the tip of the pyramid. This atom provides the
separation for the junction from the contact region which is necessary
in order to find results for the transmission independent of the
choice of $\eta$ within a large parameter window. 

Indeed, our expectations are well confirmed by the numerical data.
Fig. \ref{f6}, upper channel, shows transmission lines for
$\eta$ varying over two orders of magnitude.
All traces faithfully reproduce
the salient resonance structures in the vicinity of
1 eV about the Fermi energy, $E_{\rm F}=-5.05$ eV.
In this regime, the transmission is almost unaffected by the
change in $\eta$ even though the average DoS
changes by a factor of three, see Fig. \ref{f6}, lower panel.
Eventually, there is an impact of $\eta$ in the
tails of the resonances, where the transmission is small and
the DoS changes with $\eta$ by an order of magnitude. This regime can be better
controlled by creating an additional spacer layer of Au-atoms
between the contact atoms and the junction.

We believe, that the construction principle (\ref{e24})
is well suited to model the line broadening of resonances.
There is, however, no prediction as for the line shift, of course.
Line shifts occur, e.~g. when charge
reorganization takes place,
involving charge flow from one
subsystem to another.
Such processes can be reliably modelled only by
increasing the number of electrode atoms
in the calculation and monitoring the results.

\subsubsection{\label{sss:IIIB2} ${\rm S-Au}_3$ Coupling}

In this paragraph, we investigate a situation in which the S-atoms
couple to three Au-atoms rather than just one. There are two
motivations for doing so. First, a two or three site configuration of
the sulphur is energetically more favorable than the single site
coordination used in Sec. \ref{sss:IIIB1} and thus more likely to be
relevant for the understanding of experiments \cite{evers03prb}.
Second, we would like to
present a simple example for a situation, where the dwell time in
the cavity is not sufficiently long so that the parameter window has
closed, in which the transmission traces are independent
of the damping $\eta$. 

In general, one avoids coupling the S-atoms directly to the
surface layers, $\mathpzc{S}_{\rm L,R}$, because the change in the
local DoS near the contact may feed back into the
transmission.  Therefore, for the ${\rm S-Au}_3$-coupling
(Fig. \ref{f5a}) we include all Au-atoms into $\mathpzc{S}_{L,R}$
except for those three Au-atoms, that bind to $\mathrm{S}$.
After this minor modification, calculations proceed in
the same way as before. 

\begin{figure}
\includegraphics[width=1.0\linewidth]{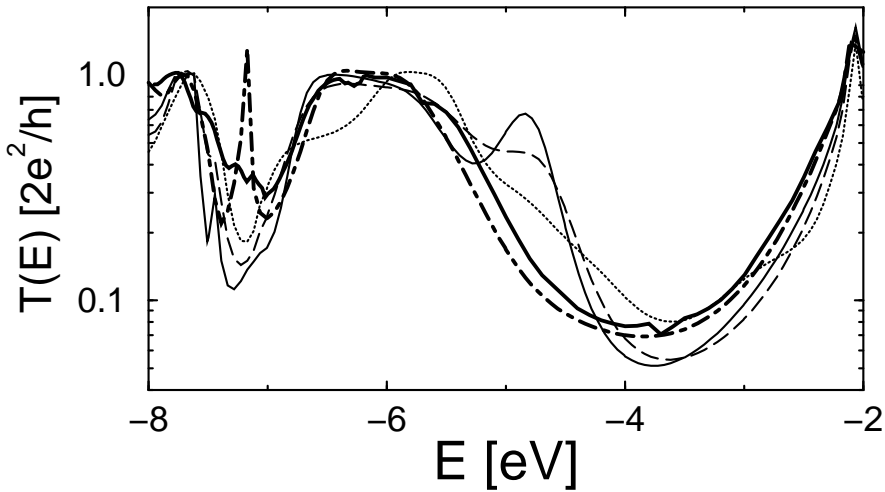}%
\caption{\label{f7}  Transmission of di-thiophenyl
  with a ${\rm S-Au}_3$ coupling (see Fig. \ref{f5a}).
   Thin traces: 14-Au pyramids with $\eta$ 0.81 eV (dotted), 2.76 eV (solid)
  and 8.1 eV (dashed). Fat traces: 55-Au 
  pyramids with $\eta$ 2.7 eV (dot-dashed) and conventional coupling
  (\ref{e24}) (solid).} 
\end{figure}

The resulting transmission is displayed in Fig. \ref{f7}.
For the 14-Au pyramids traces for three different damping
values $\eta$ are shown. They vary appreciably  one from another and
the correct result is recovered only in relatively rough terms.
The limited quality of the model coupling in the present case was to be
expected. Since the number of surface atoms that couples to the
${\rm S-Au}_3$  unit (eight) is much bigger than it was with
the  ${\rm S-Au}_1$ coupling (four), also the dwell time is
drastically reduced, too much for our simple model (\ref{e6})
to work with high precision. 

The situation is much more favorable if larger cavities are used.
Fig. \ref{f7} also shows the result for a 55-Au cluster which is in
good agreement with the conventional calculation.

\section{Discussion and Outlook}

The method relying on absorbing boundary conditions (abc),
that we have outlined in the preceeding sections,
has an important advantage over the more orthodox way
of calculating an essentially exact self energy (\ref{e1b}):
it is much easier to handle. If the abc-calculation
meets the requirements listed in Sec. (\ref{s:IIA2}),
then both methods are expected to yield identical
results for the Green's functions. 

We have demonstrated how the abc-approach can be combined with
standard quantum chemistry calculations for extended
molecules, $e\mathpzc{M}$.
Because it is not necessary any more to calculate self energies,
$\Sigma_{e\mathpzc{M}}$, transport calculations based on the
Landauer-B\"uttiker theory are greatly simplified. 

The computational effort is dominated by the quantum chemistry
calculation for $e\mathpzc{M}$.
Since the number of electrode atoms, $N_{\mathpzc{E}}$,
that needs to be included in the calculation,
is similar for both methods, one expects
that also the computational effort is roughly the same.

The number $N_{\mathpzc{E}}$ is too
large in order to allow highly correlated,
essentially exact methods to be used for calculating
$G_{e\mathpzc{M}}$.
However, calculations based on density functional
theory can be done very efficiently for these system sizes and also
Hartree-Fock calculations are within reach.
The last point
is of particular interest, because this is a prerequisite to
test functionals against each other (BP89, B3Lyp, LHF, etc.),
that have a very different degree of self interactions. 

These tests are important for the recent debate on the origin of
the discrepancy between theoretical and experimental results on the
conductance observed for several organic molecules
\cite{reimers03,evers03prb,kurth:035308,sai05,burke05a}.
The conductance of monatomic chains can be investigated very well
with the combination of methods, density functional theory (DFT) and
Landauer-B{\"u}ttiker formula, that have been employed in
Sec. \ref{s:IV} \cite{agrait03}.
By contrast, theoretical expectation for the transmission of organic
molecules tend to deviate by one or more orders of
magnitude from the experimental findings \cite{evers03prb}.

Apart from experimental difficulties also
approximations, that are implicit to the DFT based transport
formalism, could very well be responsible for this discrepancy.
This is, roughly speaking, because the Green's functions,
that derive from the (ground state)
Kohn-Sham-formalism do not necessarily provide  a good description of
the system dynamics. This description can indeed be acceptable,
if electron density is
relatively smooth and the system is at least close to metallic,
like it is in single atom metal chains. Under these conditions,
single particle wavefunctions, which are essentially plane wave
states, give a good representation of the spatial properties of the
true Green's function. Then, it is mostly the (local) spectral
properties (i.e. the band structure of the atomic wire)
that determine the transport characteristics and those can
be given quite accurately by present days DFT calculations. 

However, organic molecules are a different case, because at least in
the vicinity of the contact region, the electron density is far from
homogenous. In these regions, the spatial structure of the KS-Green's
function will in general not be a faithful representation of reality,
certainly not within local or semi-local approximations of the
exchange correlation functional. Since these very regions form exactly
the bottleneck for the transport current, theoretical calculations
employing such functionals cannot necessarily be expected to be very
precise. One can hope, that the useage of non-local functionals
will improve upon this situation. 
Which of the various non-local terms, that exist in the exact
(quasi-static non-equilibrium) functional, is the most important one and
how to implement it in practical transport calculations,
these are at present two of the most
thrilling issues in the field of molecular electronics.


\section{Acknowledgments}
The authors thank K. Busch for drawing their attention to
current applications of absorbing boundary conditions
in the field of quantum optics. Also, they
are indebted to F. Weigend for his help in using the
program package TURBOMOLE and numerous instructive discussions.
Finally, they express their gratitude to P. W\"olfle
for helpful comments on the manuscript and in particular
for his continuous support of this work. Support has also been
received from  the
{\em Center for Functional Nanostructures} at U Karlsruhe and is
gratefully acknowledged. 

\newpage

\appendix
\section{\label{a1} A Triple Barrier in a Tight Binding Wire}
A one dimensional tight binding chain with only nearest neighbor
hopping and three barriers is considered. 
\medskip

\noindent\includegraphics[width=\linewidth]{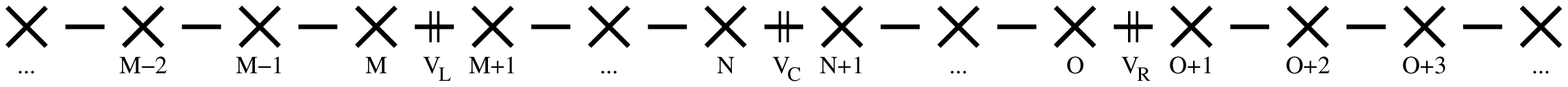}

The Hamiltonian of this problem is:
\begin{equation}
  H_{i, j} = t_{i} \left(\delta_{i, j+1} + \delta_{i+1, j}\right)  
\end{equation}
with hopping amplitudes $t_i$ given by
\begin{center}
  \begin{tabular}{rcp{2cm}l}
    $t_M$ &=& $V_L$ & left barrier   \\
    $t_N$ &=& $V_C$ & center barrier \\
    $t_O$ &=& $V_R$ & right barrier  \\
    $t_i$ &=& $1$   & everywhere else
  \end{tabular}
\end{center}
and  barriers located at positions $M$, $N$, $O$. 
The scattering states  in the four different
sections of the wire can be written as
\begin{center}
  \begin{tabular}{rcp{3cm}l}
    $\Psi_j$ &=& $e^{ikj} + re^{-ikj}$  & left lead              \\
    $\Psi_j$ &=& $ae^{ikj} + be^{-ikj}$ & left to center barrier \\
    $\Psi_j$ &=& $ce^{ikj} + de^{-ikj}$ & center to right barrier\\
    $\Psi_j$ &=& $te^{ikj}$             & right lead
  \end{tabular}
\end{center}
with energy $E(k) = -2 \cos k$
(length scales are measured with respect to $a$).

\noindent
Let us now write down the consequences of the
Schr\"odinger equation left and right of each
barrier:
\begin{eqnarray*}
  E\Psi_M &=& \Psi_{M-1} + V_L\Psi_{M+1}\\
  E(\ej{M} + r\ej{-M}) &=&\ej{(M-1)} + r\ej{(1-m)} + V_L (a\ej{(M+1)}+b\ej{-(M+1)})\\
  E\Psi_{M+1} &=& V_L\Psi_{M} + \Psi_{M+2}\\
  E(a\ej{(M+1)}+b\ej{-(M+1)})&=&V_L(\ej{M} + r\ej{-M})+a\ej{(M+2)}+b\ej{-(m+2)}\\
  E\Psi_N &=& \Psi_{N-1} + V_C\Psi_{N+1}\\
  E(a\ej{N} + b\ej{-N}) &=&a\ej{(N-1)} + b\ej{(1-N)} + V_C (c\ej{(N+1)}+d\ej{-(N+1)})\\
  E\Psi_{N+1} &=& V_C \Psi_{N} + \Psi_{N+2}\\
  E(c\ej{(N+1)}+d\ej{-(N+1)})&=&V_C(a\ej{N} + b\ej{-N})+c\ej{(N+2)}+d\ej{-(N+2)}\\
  E\Psi_O &=& \Psi_{O-1} + V_R\Psi_{O+1}\\
  E(c\ej{O} + d\ej{-O}) &=&c\ej{(O-1)} + d\ej{(1-O)} + V_R t\ej{(O+1)}\\
  E\Psi_{O+1} &=& V_R \Psi_{O} + \Psi_{O+2}\\
  Et\ej{(O+1)}&=&V_R(c\ej{O} + d\ej{-O})+t\ej{(O+2)}\\
\end{eqnarray*}
or in matrix notation:
\begin{eqnarray*}
  \begin{pmatrix}
    \ej{(M+1)} & \ej{-(M+1)}\\
    V_L\ej{M} & V_L\ej{-M}
  \end{pmatrix}
  \begin{pmatrix}
    1\\r
  \end{pmatrix} &=&
  \begin{pmatrix}
    V_L\ej{(M+1)} & V_L\ej{-(M+1)}\\
    \ej{M} & \ej{-M}
  \end{pmatrix}
  \begin{pmatrix}
    a\\b
  \end{pmatrix}\\
  \begin{pmatrix}
    \ej{(N+1)} & \ej{-(N+1)}\\
    V_C\ej{N} & V_C\ej{-N}
  \end{pmatrix}
  \begin{pmatrix}
    a\\b
  \end{pmatrix} &=&
  \begin{pmatrix}
    V_C\ej{(N+1)} & V_C\ej{-(N+1)}\\
    \ej{N} & \ej{-N}
  \end{pmatrix}
  \begin{pmatrix}
    c\\d
  \end{pmatrix}\\
  \begin{pmatrix}
    \ej{(O+1)} & \ej{-(O+1)}\\
    V_R\ej{O} & V_R \ej{-O}
  \end{pmatrix}
  \begin{pmatrix}
    c\\d
  \end{pmatrix}
  &=&
  t \ej{(O+1)}
  \begin{pmatrix}
    V_R\\
    E-\ej{}
  \end{pmatrix}.
\end{eqnarray*}
Since, our interest is in the transmission, $T = |t|^2$, we only extract an
equation for the transmission coefficient, $t$:
\begin{eqnarray}
  && \Big[({V_L}^2\ej{} - \ej{-})({V_C}^2\ej{} - \ej{-})\notag\\
  && - ({V_L}^2\ej{-(2M+1)}-\ej{-(2M+1)})
  ({V_C}^2\ej{(2N+1)} - \ej{(2N+1)})\Big]
  ({V_R}^2\ej{} - \ej{-})\notag\\
  &-&\Big[({V_L}^2\ej{} - \ej{-})({V_C}^2\ej{-(2N+1)} - \ej{-(2N+1)})\notag\\
  &&  - ({V_L}^2\ej{-(2M+1)} - \ej{-(2M+1)})({V_C}^2\ej{-} - \ej{})\Big]
  ({V_R}^2\ej{(2O+1)} - \ej{(2O+1)})\notag\\
  &=& \frac{8}{t} V_L V_C V_R \sin^3k.
  \label{appendix}
\end{eqnarray}
After specializing to the case of a strong, symmetric barrier, where $V_L {=}
V_C {=} V_R$ ${=} V \ll 1$ and $M{=}0$, $N{=}2$, $O{=}4$,
we conclude 
  \be
  \label{ea7}
  T(k) = \left( 2V^3
    \frac{\sin^3(ka)}{|\sin^2(3ka)-V^2\sin^2(2ka)+2V^2e^{ika}\sin(3ka)\sin(2ka) + \mathpzc{O}(V^4)|}\right)^2
\ee
which is also displayed in Fig. \ref{f2}, upper panel, in the body of the
paper, where also a brief discussion of this result may be found.

\printindex
\end{document}